\documentclass[english]{article}
\usepackage[letterpaper, margin=1in]{geometry}
\usepackage{float}
\usepackage[font=small,labelfont=bf]{caption}
\usepackage{amsmath}
\numberwithin{equation}{section}
\usepackage{amsmath,latexsym}
\usepackage{amssymb}
\usepackage{graphicx}
\usepackage{mathtools}
\usepackage{multicol}
\usepackage{subcaption}
\usepackage[T1]{fontenc}
\usepackage[latin9]{luainputenc}
\usepackage{babel}
\pdfoutput=1
\usepackage{hyperref}
\usepackage{pdfpages}
\usepackage{microtype}

\begin{document}

\title{The Sturdiness of the Shell Model: Informal Review}
\author{Castaly Fan and Larry Zamick\\
\vspace{1em}\\
Department of Physics and Astronomy, Rutgers University, \\
 Piscataway, New Jersey 08854, USA}

\maketitle

\begin{abstract}
    With shell model codes being able to encompassing larger and larger spaces we find that the percentage occupancy of the leading spaces becomes smaller and smaller. How can the shell model survive in such circumstances? We will not solve this puzzle here but rather will show examples where, with some explanations, the shell model holds fast. We will use nuclear moments as an example.
\end{abstract}

\section{Introduction}
    With $n$ nucleons of one kind there are simple formulas for nuclear moments in a single $j$ shell. For example all $g$ factors should be the same. From this it follows that for states of the same $J$ the magnetic moments should be the same. The magnetic moment of a free neutron (in units of nuclear magnetons) is $\mu_{n} = -1.913$ and that of a free proton is $\mu_{p} =+2.793$. In a single $j$ shell of neutrons with $j=l +1/2$ the magnetic moments are predicted to be the same as those of a free neutron-namely $-1.913$ ; for protons it is $(2.793 +l) (\mu_{n})$. Here $L$ is the orbital angular momentum.
    
    The single particle magnetic moments ,commonly called the Schmidt moments are given here: 
    \begin{enumerate}
        \item for an odd proton: 
        \begin{itemize}
            \item $\mu = j-1/2 + \mu_{p}$ for $j = l + 1/2$ 
            \item $\mu = j /(j + 1) [ j + 3/2 - \mu_{p} ]$ for $j = l - 1/2$ 
        \end{itemize}
        \item for an odd neutron:
        \begin{itemize}
            \item $\mu = \mu_{n}$ for $j = l + 1/2$
            \item $\mu = - j/ ( j+1) \mu_{n}$ for $j =l-1/2$. 
        \end{itemize}
    \end{enumerate}    
    We can discuss these in a more physical manner. The magnetic moment of a free neutron (in units of nuclear magnetons ) is $\mu_{n} = -1.913$ and that of a free proton is $\mu_{p} =+2.793$. In a single $j$ shell of $n$ neutrons with $j=l +1/2$ the magnetic moments are predicted to be the same as those of a free neutron-namely $-1.913$; for $n$ protons it is $(2.793 +l)$. Here $l$ is the orbital angular momentum. For a $j=l-1/2$ neutron we have a quantum effect so that the magnetic moment is only minus that of a free neutron in the large $j$ limit. In general it is $-j / (j+1)$ that of a free neutron.
    
    For quadrupole moments there is also an $n$ dependent simple formula for ground states of odd nuclei in a single $j$ shell
        \begin{equation*}
            Q(n) =[(2j+1-2n)/(2j-1)] Q(sp)
        \end{equation*}
    Note that for a single hole $n=2 j$. The formula becomes $Q= - Q(sp)$. I.e. the quadruple moment of a hole is minus that of a particle. We can understand this another way. A nuclear moment is the expectation value of a moment operator in a state with $M=J$, 
        \begin{equation*}
             Q^{2}= \langle \Psi^{J}_{J}| Q^{2}_{0} | \Psi^{J}_{J} \rangle.
        \end{equation*}
    To create a hole nucleus in a state with $M=J$ we have to remove a nucleon from a closed shell with $M=-J$. The value of $Q$ for a closed shell is zero-this is the the sum of $Q$ for the hole nucleus and the nucleon removed. The value of $Q^2$ in a state with $M=J$ is the same as it is for $M= -J$ -- namely $Q(sp)$. So we have $Q(\text{hole})+Q(sp)=0$ or $Q(\text{hole})=-Q(sp)$. For magnets moments we have the opposite the value for $-J$ is minus that for $+J$. Thus we have 2 minus signs and $\mu(\text{hole})=\mu(sp)$.

\section{Examples}

\subsection{Charge quadruple moments}

    As a first example of the sturdiness of the shell model we look at the work of Ruiz et al. \cite{1} on measurements and theoretical analysis quadrupole moments of odd A nuclei in the ``f-p'' region. They measured the quadruple moments of the $J=7/2^{-}$ ground states of Calcium isotopes with $A=43,45,47$ which have ground state spins $J=7/2^{-}$. They did not do $A=41$ but this case could be obtained from another source. They also obtained results for $A=49,51$ with $J=3/2^{-}$ spins.

    A starting point for $A=41$ to 47 would be the f$_{7/2}$ shell while for $A=49,51$ it would be the p$_{3/2}$ shell. The theoretical calculations were performs with many interactions and different model spaces. The latter include complete pf space, (pf$+$g$_{9/2}$) and breaking the $^{40}$Ca core by allowing 2p-2h admixtures. They use effective charges of 1.5 for the protons and 0.5 for the neutrons. In general the calculations are in excellent agreement with the measurements. We will not go into further details about the calculations except to say that they involve an enormous number of shell model configurations.

\begin{figure}[H]
        \centering
        \captionsetup{width=0.8\linewidth}
        \includegraphics[width=0.65\textwidth]{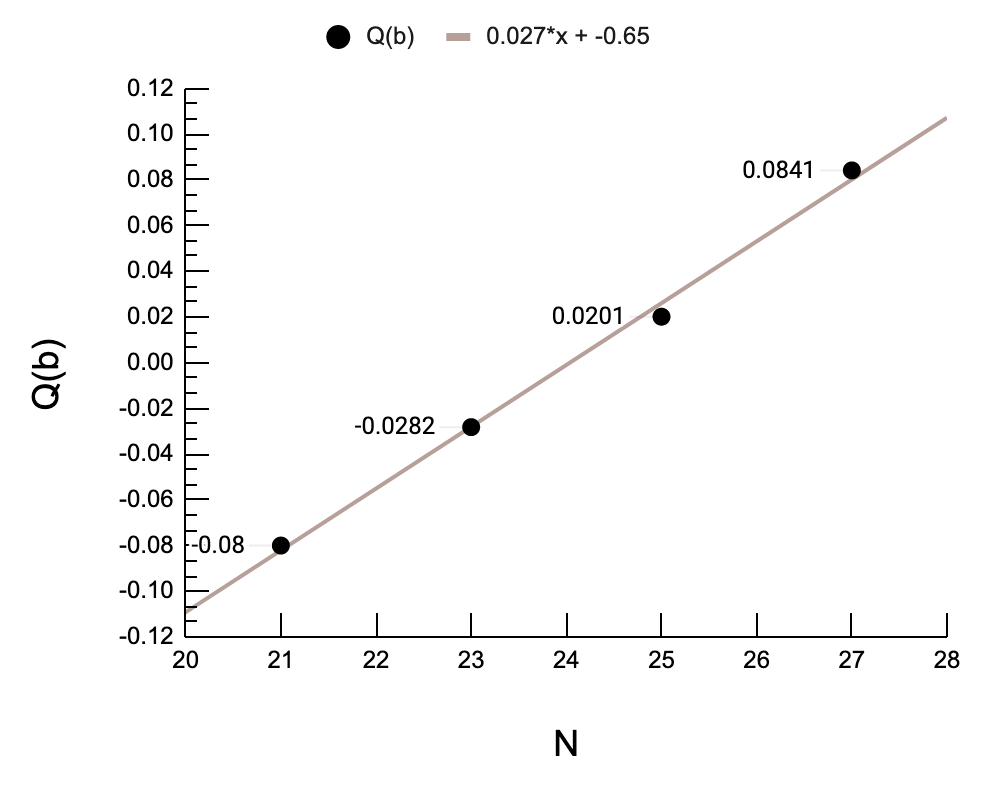}
        \caption{The quadrupole moments vs. $A$ from $A = 41$ to 47.}
        \label{fig1}
    \end{figure}

    Rather in Fig \ref{fig1} we show the quadrupole moments vs. $A$ and show the the remarkable result that the measured moments from $A=41$ to $A=47$ lie, to an excellent approximation on a straight line. As noted in the introduction this is exactly what a single $j$ calculation predicts. To repeat $Q= (2j-1-2n)/(2j-1)*Q(s.p.)$ This simple result seems to survive the large shell attack. For $A=51$ the measured quadrupole moment $Q=+0.04$ b. It is nearly equal and opposite of that for $A=49$ $Q=-0.04$ b. This is the prediction of the simplest shell model in which $A=49$ consist of a single p$_{3/2}$ neuron and $A= 51$ of a p$_{3/2}$ hole.

    Before leaving this section we should mention that a purist might say that the real prediction of single $j$ is that all the charge quadrupole moments are zero because the neutrons have no charge. We have to assign an effective charge to the neutrons, popular choice being $e_{\text{eff}}=0.5$. But note that even the large space calculations including those of \cite{1} require effective charges in order to get agreement with experiment. In first order perturbation theory the effective charge comes from $\Delta N =2$ excitations. For example for $^{41}$Ca excitations from 0p to 1p; from 0d to 0g, 1d, and 2s. As large as model spaces are in \cite{1} and in nearly all other calculations these configurations are not present and one needs to insert effective charges.

    \subsection{Magnetic moments}

\begin{figure}[H]
        \centering
        \captionsetup{width=0.8\linewidth}
        \includegraphics[width=0.65\textwidth]{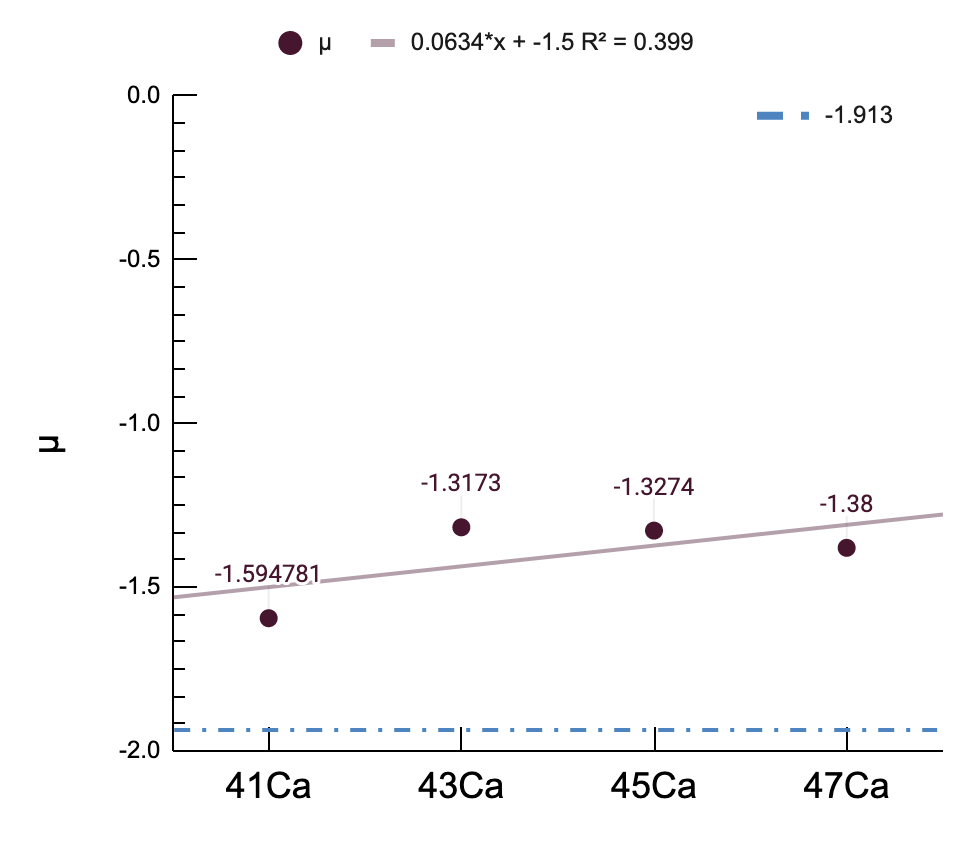}
        \caption{The Ca isotopes from $A = 41$ to 47 based on Ruiz et al's results \cite{1}.}
        \label{fig2}
    \end{figure}

    Let us next look at magnetic moments. In Fig \ref{fig2} we show these for the Ca isotopes ($A=41$) to 47 again from the work of Ruiz et al. \cite{1}. There is considerable deviation from the single $j$ value $\mu$ (free neutron) = $-1.913\mu_{n}$. They are closer to $-1.4$. This is called quenching. But the prediction of the single $j$ that all $g$ factors should be the same, is approximately realized. Of course the deviations from the simple result is of great interest and should be addressed.
    
\begin{figure}[H]
        \centering
        \captionsetup{width=0.8\linewidth}
        \includegraphics[width=0.7\textwidth]{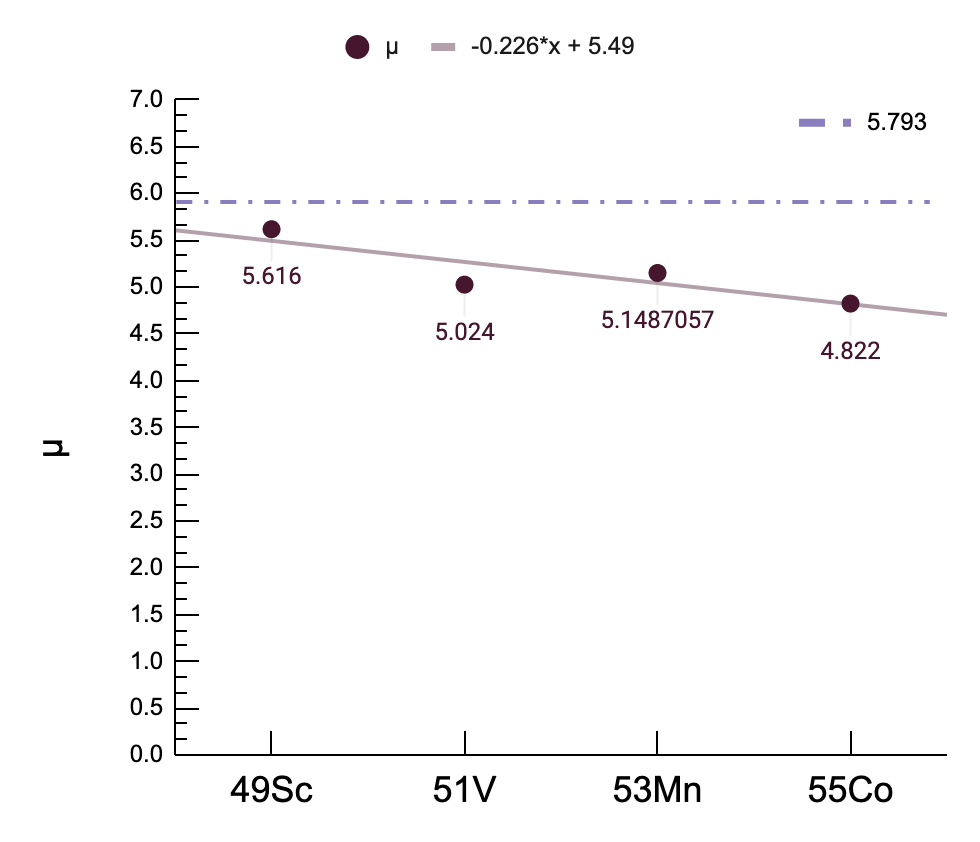}
        \caption{The $N=28$ isotones.}
        \label{fig3}
    \end{figure}

    Let us next look at the $N=28$ isotones in Fig \ref{fig3}. The experiments were performed by Speidel et al. \cite{2}. The nuclei involved are $^{49}$Sc, $^{51}$V, $^{53}$Mn, and $^{55}$Co. The simplest configuration consist of f$_{7/2}$ protons. The Schmidt value is $\mu$ (free proton) +3 = 5.793 $\mu_{n}$. Note that there is a bit of quenching but the most striking thing is that the magnetic moments lie on a straight line with a negative slope.

    Now we will reluctantly introduce a bit of theory. For a closed major shell plus on nucleon there are no corrections to the magnetic moment s in first order perturbation theory. One has to go to second order as has been done by Ichimura and Yazaki \cite{3} and Mavromatis, Zamick and Brown \cite{4}. If one looks hard at Fig \ref{fig2} we see that the ``one nucleon'' magnetic moments ($^{41}$Ca and $^{49}$Sc) are a bit off from the others. They are a bit closer to Schmidt. On the other hand if one has an open shell of valence nucleons one can get first order calculations. Such calculations were performed early on by Arima and Horie \cite{5}. Many people remember that these calculations provide a microscopic explanation of quenching but they have perhaps forgotten that the quenching is $n$ dependent. Arima and Horie's prediction \cite{5} that $g$ factors will lie on a straight line with a slope is beautifully realized by the $N=28$ isotones shown in Fig \ref{fig3}.

    We should not be too complacent. There are cases where there is a breakdown of the simple shell model. Consider the measured g factors of 2+ states in $^{42}$Ca and $^{44}$Ca. The $g$ factors are close to zero. In the simple single $j$ shell model (f$_{7/2}$ neutron) the $g$ factor is $-1.913/3.5 = -0.547$. A simple explanation, a la Gerace and Green \cite{5} is that there is a low lying highly deformed $J=2^{+}$ state with a $g$ factor $Z/A$ which is close to $+0.5$. So simple math $(0.5 -0.5)/2 =0$. People are having trouble getting these highly deformed states low enough in energy, even in very large shell model calculations.
    
    Even in very late shell model calculations. However Zheng, Zamick and Berdichevsky \cite{21}, in a work entitled ``Calculations of many-particle - many-hole deformed state energies: Near degeneracies, deformation condensates'', were able to obtain these low lying intruder states in deformed Hartree-Fock calculations with Skyrme interactions.
    
    \begin{figure}[H]
        \centering
        \captionsetup{width=0.65\linewidth}
        \includegraphics[width=0.6\textwidth]{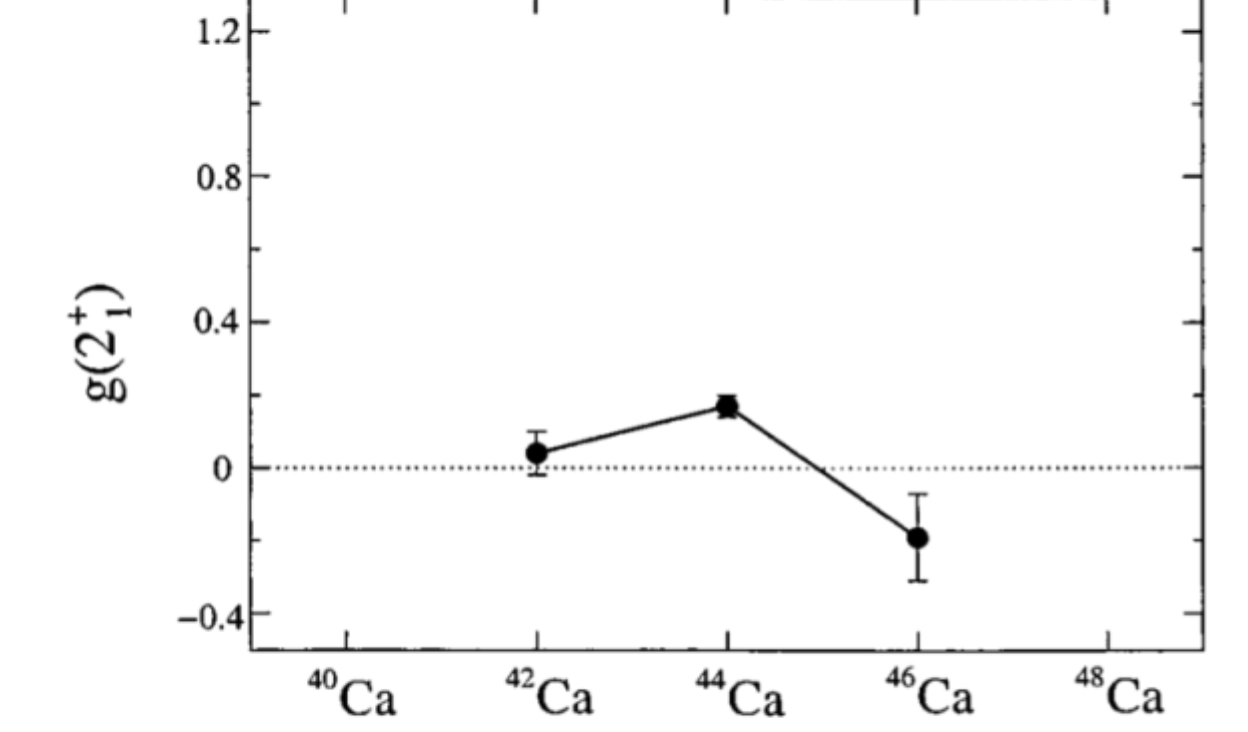}
        \caption{The $g$ factors of lowest $2^+$ states in even-even Ca isotopes.}
        \label{fig4}
    \end{figure}

\begin{figure}[H]
        \centering
        \captionsetup{width=0.75\linewidth}
        \includegraphics[width=0.7\textwidth]{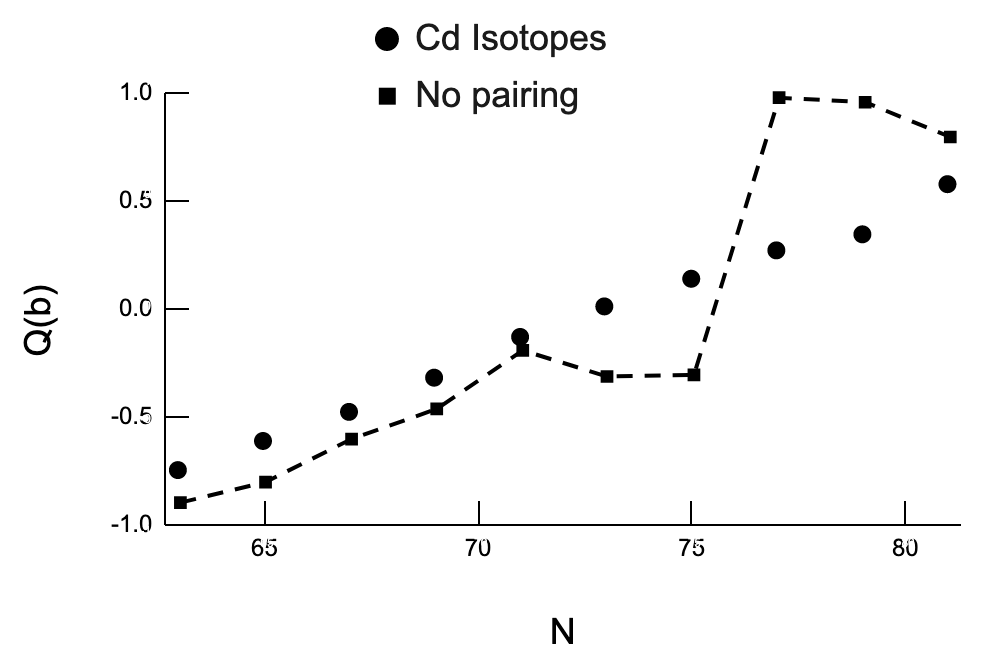}
        \caption{Quadrupole Moments of $J=11/2^{-}$ states in odd $N$ Cd isotopes. The closed circles correspond approximately both to experimental results of \cite{12} and CFTD plus pairing theory of ref. \cite{14}. The dashes line is theory of ref. \cite{14} without pairing.}
        \label{fig5}
    \end{figure}

    Lastly in this section we discuss the strange case of quadrupole moments of $J=11/2^{-}$ states in the odd $N$ cadmium isotopes from $A=111$ to $129$. We show in Fig \ref{fig5}, which redrawn from the work of D.T. Yordanov et al. \cite{12}\cite{13}. 
    
    At first thought the fact that these also lie very closely on a straight line might be considered a vindication of the simple shell model. However a simple count tells us there are too many entries in the figure. In the single $j$ shell There should only be 6 nuclei with $J=11/2^{-}$ ground states corresponding to 1, 3, 5, 7, 9, and 11 neutrons int the h$_{11/2}$ shell.

    First one should look at the experimental situation more carefully.e.g. on the nndc website \cite{15}. While for $A=129$ $J=11/2^{-}$ is the ground state for all other isotopes we have $J=11/2^{-}$ as excited $i$ states. For example in $^{111}$Cd the order of levels is $1/2^{+}$, $5/2^{+}$, $3/2^{+}$ and $11/2^{-}$ with energies, 0, 245, 342 and 396 keV respectively. The $11/2^{-}$ state cannot decay by M1, E1 or E2 to the lower states and so it is long lived i.e. isomeric. The long life makes it possible for the quadrupole moment to be measured. This problem has been addressed theoretically by P. W. Zhao, S. Q. Zhang and J. Meng \cite{14}. They use an approach which they call covariant density functional theory (CDFT). We will not go into too much detail here but it is closely related to the Hartree Fock theory where one solves for an intrinsic state from which one either projects out states of good angular momentum or uses a Bohr-Mottelson formula which relates a laboratory quadruple moment with the intrinsic one.
    
    But that is not enough. One has to add paring correlations which smear out the occupations numbers for various orbits including h$_{11/2}$. The pairing clears out abrupt changes and yields a simple linear behavior. So here we have an example where a complex theory is required to produce a simple result.
    
    Many other examples of the systematics of magnetic and quadrupole moments are contained in the  beautiful article by Gerda Nyens \cite{18}. 

\section{Isotope shifts}

\begin{figure}[H]
        \centering
        \captionsetup{width=0.75\linewidth}
        \includegraphics[width=0.7\textwidth]{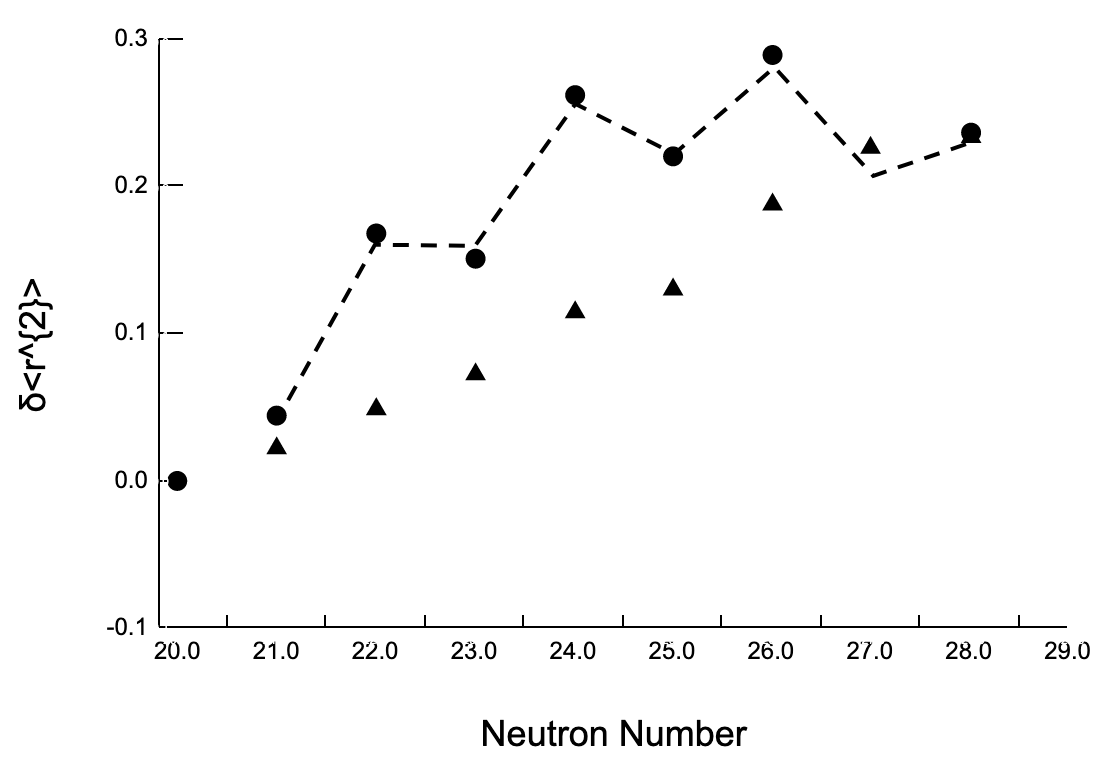}
        \caption{Isotope shifts in the odd $N$ Ar Isotopes. The filled circles correspond to the experiment of Blaum et al. \cite{9} and the dashed line the Zamick formula \cite{10}. The triangles correspond to spherical Hartee-Fock calculation \cite{19}\cite{20}}
        \label{fig6}
    \end{figure}

    In Fig \ref{fig6} we show measured values of isotope shifts in the Argon Isotopes by Blau et al \cite{9} (open circles). Also shown are spherical Hartree-Fock calculations in closed triangles [[HHH]], as well as a formula by Zamick \cite{10} and by Talmi \cite{19}\cite{20} which will soon be discussed.

    Note that the data shows a lot of even-odd staggering but the HF calculations do not. The Zamick-Talmi calculations yield excellent fits to the data and have the even-odd scattering features  well under control. In order to get the even-odd staggering Zamick \cite{10} introduced a 2 body effective radius operator in addition to the one body term.
    
    We simply make the assumption that the effective charge radius operator has a two-body part as well as one body part
        \begin{equation}
            \delta r_{\text{eff}}^{2} = \sum_{i}O(i) + \sum_{i<j}V(i,j)
        \end{equation}
     where the symbol $V$ for the two-body part has been written to suggest the similarity with the two-body potential, since both are scalars.
    
    The problem of evaluating this operator for n particles in the $j = f_{7/2}$ shell is exactly the same problem as calculating the binding energies of nuclei whose configuration consists of several nucleons in a single j shell. This problem has been solved and used with great success by the ``Israeli group'' including de-Shalit, Racah, Talmi, Thieberger, and Unna \cite{11}. In analogy with their binding energy formula we get for the change in charge radius
        \begin{equation}
            \delta r^{2} (40 + n) = nC + \frac{n(n-1)}{2} \alpha + \left[ \frac{n}{2} \right]\beta,
        \end{equation}
    where
        \begin{equation}
            \left[ \frac{n}{2} \right] = 
            \begin{dcases}
                \frac{n}{2} & \quad \text{for even $n$}\\
                \frac{n-1}{2} & \quad \text{for odd $n$}.
            \end{dcases}
        \end{equation}
    The parameter $C$ comes from the one-body part and is equal to $\delta r^{2}(41)$, the difference in charge radius of $^{41}$Ca and $^{40}$Ca. The quantities $\alpha$ and $\beta$ come from the two-body part
        \begin{equation}
        \begin{split}
            \alpha &= -\frac{2(j+1)\bar{E}_{2} - E_{0}}{2j + 1},\\
            \beta &= \frac{2(j+1)(\bar{E}_{2} - E_{0})}{2j + 1},
        \end{split}
        \end{equation}
    where
        \begin{equation}
        \begin{split}
            E_{0} &= \langle j^{2}J = 0 | V | j^{2}J = 0 \rangle\\
            \bar{E}_{2} &= \frac{\sum_{J\neq 0}(2J+1) \langle j^{2}J | V | j^{2}J\rangle}{\sum_{J\neq 0}(2J+1)}.
        \end{split}
        \end{equation}

\section{Two body interaction}

    If one has a closed shell plus 2 valence nucleons in a single $j$ shell, then with only a single particle spherical potential there is a great deal of degeneracy. Let us use the f$_{7/2}$ shell as an example. With a valence neuron ad valence proton the degeneracy is $(2j+1)(2j+1) =64$. We can remove some of the degeneracy by introducing a residual 2 particle interaction $V$. In first order perturbation theory one gets a spectrum as follows: If $V$ is rotationally invariant we get states classified by the total angular momentum of the 2 nucleons $J$. States with energy $E(J)$ still has a $(2J+1)$ degeneracy with $M_J$ ranging from $-J$ to $J$.
    
    We take the matrix elements $E(J)$ from experiment, i.e. from the spectrum of $^{42}$Sc. We make the association $E(J)$ is the experimental excitation energy in 42Sc of the lowest state of angular momentum $J$. In first order perturbation theory we have $\langle [f_{7/2} f_{7/2} ] |V| [f_{7/2} f_{7/2} ]\rangle ^{J} =E^{*}(J)$. (We often shift things so that the $J=0$ ground state is at zero energy.That will not affect the spectrum). For example we make the association that  $\langle [f_{7/2} f_{7/2} ]V| [f_{7/2} f_{7/2} ]\rangle^{2} = 1.580$ MeV with MBZE(2006) Note that we do not deal directly with $V$ which can be very complicated but rather with a matrix element of $V$. This procedure is a very common one - it is even used in multi-shell calculations. In Tables \ref{tab1} and \ref{tab2} we list the empirical 2 body matrix elects that have bee used in the f$_{7/2}$ and g$_{9/2}$ regions. In Table \ref{tab1} we show various sets of two particle matrix elements for the f$_{7/2}$ region. The one MBZ(1964) was obtained where the empirical data was sparse and is inferior the MBZE(2006). We also show matrix events obtained from the spectra of 2 hole ($^{54}$Co). In the ideal case the spectra of 2 holes should be the same as that of 2 particles, but due to configuration mixing that is not exactly the case. We throw in the Q.Q interaction matrix elements which roughly are similar to the empirical ones. In the g$_{9/2}$ shell Table \ref{tab2} we only work close to $^{100}$Sn . In the other limit the shell model breaks down because $^{80}$Zr is strongly deformed.

    \begin{table}[H]
        \centering
        \captionsetup{width=0.75\linewidth}
        \caption{Two body matrix elements used in the f$_{7/2}$ shell.}
        {\renewcommand{\arraystretch}{1.5}
        \begin{tabular}{|l|l|l|l|l|l|l|l|l|}
            \hline
            $J$ & 0 & 1 & 2 & 3 & 4 & 5 & 6 & 7 \\ \hline
            MBZ(1964) & 0.0000 & 1.036 & 1.509 & 2.248 & 2.998 & 1.958 & 3.400 & 0.617 \\ \hline
            MBZE (2006)$^{42}$Sc & 0.0000 & 0.6110 & 1.5803 & 1.4904 & 2.8153 & 1.5100 & 3.2420 & 0.6163\\ \hline
            hole-hole $^{54}$Cr & 0.0000 & 0.9369 & 1.4457 & 1.8215 & 2.6450 & 1.8770 & 2.9000 & 0.1974 \\ \hline
            f$_{7/2}$ Q.Q & 0.0000 & 0.3655 & 1.0232 & 1.8270 & 2.5579 & 2.9233 & 2.5580 & 1.0232 \\ \hline
            \end{tabular}}
        \label{tab1}
    \end{table}
    
    \begin{table}[H]
        \centering
        \captionsetup{width=0.75\linewidth}
        \caption{Two body matrix elements used in the g$_{9/2}$ shell.}
        {\renewcommand{\arraystretch}{1.5}
        \begin{tabular}{|l|l|l|l|l|l|l|l|l|l|l|}
            \hline
            $J$ & 0 & 1 & 2 & 3 & 4 & 5 & 6 & 7 & 8 & 9 \\ \hline
            Qi et al. & 0.000 & 1.220 & 1.458 & 1.592 & 2.283 & 1.882 & 2.549 & 1.930 & 2.688 & 0.626\\ \hline
            CCGI & 0.000 & 0.829 & 1.710 & 1.877 & 2.217 & 2.046 & 2.383 & 1.913 & 2.527 & 0.915\\ \hline
            INTd & 0.0000 & 1.1387 & 1.3947 & 1.8230 & 2.0283 & 1.9215 & 2.2802 & 1.8797 & 2.4275 & 0.7500\\
            \hline
            g$_{9/2}$ Q.Q & 0.0000 & 0.3536 & 1.0168 & 1.8990 & 2.8736 & 3.7618 & 4.3325 & 4.3325 & 3.4483 & 1.3262 \\ \hline
        \end{tabular}}
        \label{tab2}
    \end{table}
    
    As will be discussed later if we have 2 nucleons in a single $j$ shell the states of even $J$ have isospin one and the odd $J$ isospin zero. We show in Tables \ref{tab3} and \ref{tab4} all $J$, even $J$ and odd $J$ centroids. Note that in all cases the odd $J$ centroids have a smaller value that the even ones. This means that on average the odd $J(T=0)$ interaction is more attractive than $J(T=1)$. This despite the fact that there has been much emphasis in the literature on $J=0(T=1)$ pairing. We also show standard deviations.

    The centroid of interactions is given by
        \begin{equation}
            C = \frac{\sum_{i} (2J_{i} + 1) E_{i}}{\sum_{i}(2J_{i} + 1)}
        \end{equation}
    One can write down either odd terms ($J = 1, 3, 5, ...$), even terms ($J = 0, 2, 4, ...$), or all ($J = 0, 1, 2, ...$). For instance, we have g$_{9/2}$ Q.Q in Table \ref{tab2}, and we here assign the values with $E_{i}$. Hence, we have
        \begin{equation*}
        \begin{split}
             C_{\text{Odd}} &= \frac{3\times 0.3536 + 7\times 1.8990 +  ... + 19\times 1.3262}{55} \sim 1.6648\\
             C_{\text{Even}} &= \frac{1\times 0 + 5\times 1.0168 +  ... + 17\times 3.4483}{45} \sim 2.0334\\
             C_{\text{All}} &= \frac{1\times 0 + 3\times 0.3536 + 5\times 1.0168 + ... + 19\times 1.3262}{100} \sim 1.8302
        \end{split}
        \end{equation*}
    for g$_{9/2}$ Q.Q. The results are shown in Table \ref{tab3} and Table \ref{tab4}, where the standard deviation (s.d.) is followed by
        \begin{equation}
            \sigma = \sqrt{\frac{\sum_{i}^{n} (x_{i} - \Bar{x})^{2}(2J_{i} + 1)}{n-1}}
        \end{equation}
    Besides, we introduced the deviation among the values,
        \begin{equation}
            \text{deviation} = \frac{\text{Even} - \text{Odd}}{\text{All}}\times 100\%
        \end{equation}
    In the calculations, we have normalized the values by multiplying 0.9244 with f$_{7/2}$ Q.Q and multiplying 0.6272 with g$_{9/2}$ Q.Q. The purpose of the normalization constants is to let f$_{7/2}$ Q.Q's centroid be the same as MBZE(2006)$^{42}$Sc's result (1.7735) and to let g$_{9/2}$ Q.Q's centroid be the same as Qi et al.'s (1.8302).
    
    \begin{table}[H]
        \centering
        \captionsetup{width=0.75\linewidth}
        \caption{The centroids, standard deviations, and deviations of the f$_{7/2}$ shell, with odd, even, and all cases for calculations.}
        {\renewcommand{\arraystretch}{1.5}
        \begin{tabular}{|l||ll|ll|ll|l|}
            \hline
                & Odd & & Even & & All & & Deviation \\ \cline{2-7}
                & \multicolumn{1}{l|}{centroid} & s.d.   & \multicolumn{1}{l|}{centroid} & s.d.   & \multicolumn{1}{l|}{centroid} & s.d.   &           \\ \hline\hline
            MBZ(1964) & \multicolumn{1}{l|}{1.3788}   & 0.7179 & \multicolumn{1}{l|}{2.8117}   & 0.8862 & \multicolumn{1}{l|}{2.0057}   & 1.0659 & 71.44\%   \\ \hline
            MBZE(2006) $^{42}$Sc & \multicolumn{1}{l|}{1.0589}   & 0.4498 & \multicolumn{1}{l|}{2.6923}   & 0.8050 & \multicolumn{1}{l|}{1.7735}   & 1.0282 & 92.10\%   \\ \hline
            Hole-hole $^{54}$Cr  & \multicolumn{1}{l|}{1.0880}   & 0.8032 & \multicolumn{1}{l|}{2.4548}   & 0.7206 & \multicolumn{1}{l|}{1.6860}   & 1.0237 & 81.07\%   \\ \hline
            f$_{7/2}$ Q.Q        & \multicolumn{1}{l|}{1.5764}   & 0.9163 & \multicolumn{1}{l|}{2.0268}   & 0.7540 & \multicolumn{1}{l|}{1.7735}   & 0.8770 & 25.39\%   \\ \hline
            \end{tabular}}
        \label{tab3}
    \end{table}

    \begin{table}[H]
        \centering
        \captionsetup{width=0.75\linewidth}
        \caption{The centroids, standard deviations, and deviations of the g$_{9/2}$ shell, with odd, even, and all cases for calculations.}
        {\renewcommand{\arraystretch}{1.5}
        \begin{tabular}{|l||ll|ll|ll|l|}
            \hline
              & Odd & & Even & & All & & Deviation \\ \cline{2-7}
              & \multicolumn{1}{l|}{centroid} & s.d.   & \multicolumn{1}{l|}{centroid} & s.d.   & \multicolumn{1}{l|}{centroid} & s.d.   &           \\ \hline\hline
            CCGI      & \multicolumn{1}{l|}{1.5311}   & 0.5205 & \multicolumn{1}{l|}{2.2765}   & 0.4273 & \multicolumn{1}{l|}{1.8665}   & 0.6065 & 39.94\%   \\ \hline
            Qi et al. & \multicolumn{1}{l|}{1.3882}   & 0.5853 & \multicolumn{1}{l|}{2.3704}   & 0.5172 & \multicolumn{1}{l|}{1.8302}   & 0.7371 & 53.67\%   \\ \hline
            INTd      & \multicolumn{1}{l|}{1.4502}   & 0.5407 & \multicolumn{1}{l|}{2.1364}   & 0.4555 & \multicolumn{1}{l|}{1.7590}   & 0.6078 & 39.01\%   \\ \hline
            g$_{9/2}$ Q.Q  & \multicolumn{1}{l|}{1.6648}   & 1.7365 & \multicolumn{1}{l|}{2.0334}   & 1.6404 & \multicolumn{1}{l|}{1.8302}   & 1.7078 & 20.14\%   \\ \hline
            \end{tabular}}
        \label{tab4}
    \end{table}

    \begin{table}[H]
        \centering
        \caption{The spectra of $^{42}$Ca, $^{42}$Sc, and $^{42}$Ti.}
        {\renewcommand{\arraystretch}{1.5}
        \begin{tabular}{|p{1.5cm}||p{2.5cm}|p{2.5cm}|p{2.5cm}|}
        \hline
        & $^{42}$Ca   & $^{42}$Sc   & $^{42}$Ti   \\ \hline\hline
        $J=0$ & 0.0000 & 0.0000 & 0.0000 \\ \hline
        $J=1$ &  - & 0.6110 &  - \\ \hline
        $J=2$ & 1.5247 & 1.5803 & 1.5546 \\ \hline
        $J=3$ & - & 1.4904 & -  \\ \hline
        $J=4$ & 2.7524 & 2.8153 & 2.6746 \\ \hline
        $J=5$ & -  & 1.5100 & -  \\ \hline
        $J=6$ & 3.1893 & 3.2420 & 3.0430 \\ \hline
        $J=7$ & -  & 0.6163 &  - \\ \hline
        \end{tabular}}
    \label{tab6}
    \end{table}

    \begin{figure}[H]
        \centering
        \captionsetup{width=0.65\linewidth}
        \includegraphics[width=0.6\textwidth]{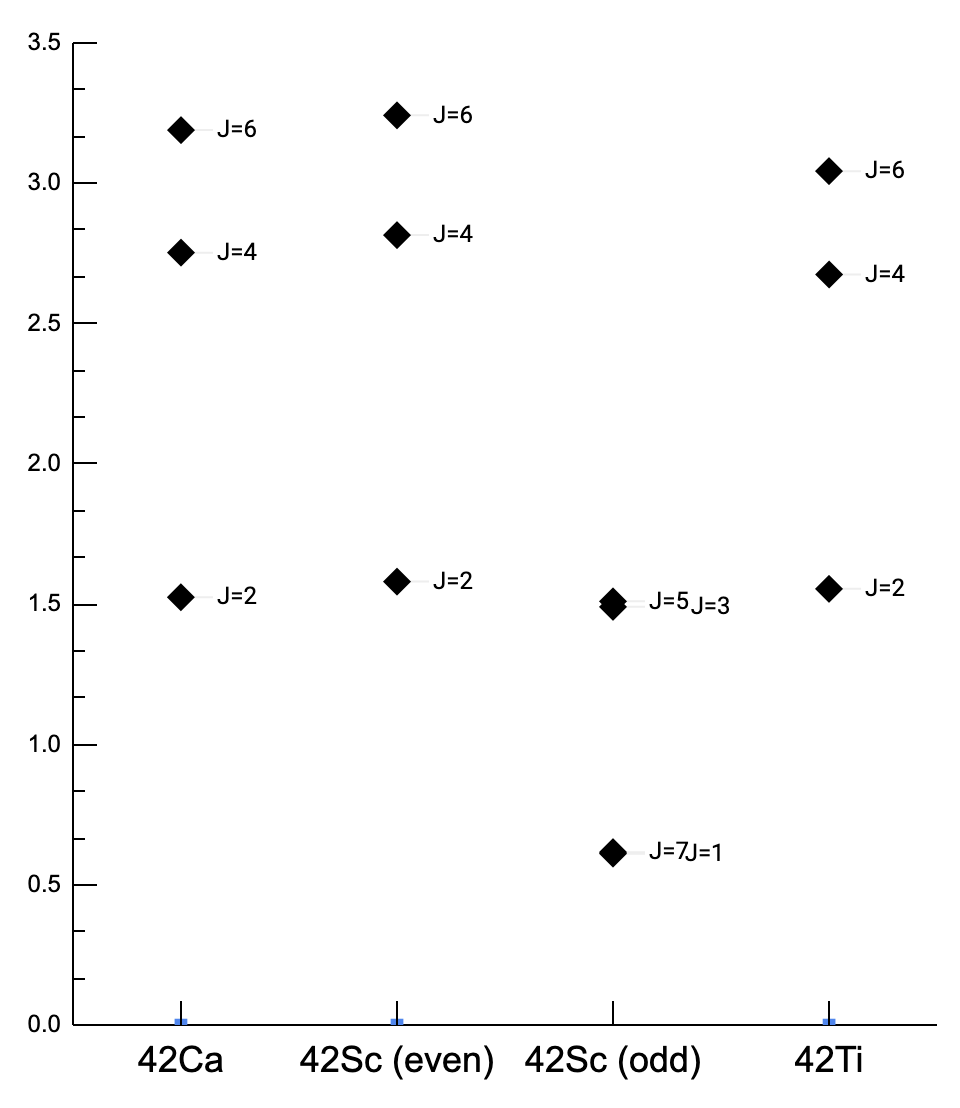}
        \caption{Energy levels of $^{42}$Ca, $^{42}$Sc and $^{42}$Ti shown in order to display the near charge independence of the nuclear force.}
        \label{fig7}
    \end{figure}

    Note in Table \ref{tab6} that the excitation energies of the even $J$ states in $^{42}$Ca, $^{42}$Sc and $^{42}$Ti are early the same. This is evidence of the charge independence of the nuclear force. The fact that odd $J$ states appear only in $^{42}$Sc shows the Pauli principle in action. In $^{42}$Ca and $^{42}$Ti we have 2 identical nucleons so we can only have antisymmetric states . This tells us that in the $j^2$ configuration states with even $J$ are antisymmetric. In $^{42}$Sc we do not have identical nucleons so we can have symmetric states. These are the odd $J$ states. And the there is the multiplicity rule. Even $J$ states occur 3 times so $(2T+1) = 3$ and so $T=1$. The odd $J$ states occur only once so $(2T+1) =1$.

    Although the matrix elements from ``experiment'' for MBZ(1964) are not as good as the ones of MBZE(2006) we include them to make a point. This has to do with isomeric states in $^{43}$Sc. In Table \ref{tab7} we show the energies of high lying states in $^{43}$Sc with the 2 interactions.
    
    \begin{table}[H]
        \centering
        \caption{Calculated Energies of high lying states in $^{43}$Sc.}
        {\renewcommand{\arraystretch}{1.5}
        \begin{tabular}{|l||l|l|l|}
        \hline
        J    & MBZ(1964) & MBZE (2006) & experiment\\ \hline\hline
        15/2 & 3.71      & 3.51        & 2.99  \\     \hline
        17/2 & 4.62      & 4.30        & 4.38   \\    \hline
        19/2 & 3.64      & 3.64        & 3.12 \\ \hline      
        \end{tabular}}
    \label{tab7}
    \end{table}
    
    We see, and this is discussed in Talmi's book on page 853 \cite{16} that with MBZ(1964) the $19/2^{-}$ state is lower In energy than $15/2^{-}$ and so we have the prediction of a spin gap isomer which would be very long lived. However with MBZE(1964). The $19/2^{-}$ state is above $15/2^{-}$ We still have an isomer because of the small entry difference between the 2 states but it is a weaker isomer. MBZE(2006) agrees with experiment - $J= 19/2^{-}$ is a bit higher than $15/2^{-}$. For additional work on isomers please see the work of P.C. Srivastava and L. Zamick \cite{17}.

\end{document}